\documentclass[article,nojss]{jss}
\usepackage{amsmath}
\usepackage{amssymb}
\newcommand{\M}[1]{\mathbf{#1}}
\newcommand{\R}{\mathbb{R}}

\author{Kasper Kristensen\\DTU Compute \And 
        Anders Nielsen\\DTU Aqua \And 
        Casper W. Berg\\DTU Aqua \AND
        Hans Skaug\\University of Bergen \And
        Brad Bell\\IHME UW
}
\title{\pkg{TMB}: Automatic Differentiation and Laplace Approximation} 

\Plainauthor{Kasper Kristensen, Anders Nielsen, Casper W. Berg, Hans Skaug, Brad Bell}
\Plaintitle{TMB: Automatic Differentiation and Laplace Approximation} 
\Shorttitle{\pkg{TMB}: Template Model Builder}

\Abstract{
\pkg{TMB} is an open source \proglang{R} package that enables quick
implementation of complex nonlinear random effect (latent variable)
models in a manner similar to the established AD Model Builder package
(\pkg{ADMB}, admb-project.org) \citep{ADModelBuilder2011}.  In
addition, it offers easy access to parallel computations.  The user
defines the joint likelihood for the data and the random effects as a
\proglang{C++} template function, while all the other operations are
done in \proglang{R}; e.g., reading in the data.  The package
evaluates and maximizes the Laplace approximation of the marginal
likelihood where the random effects are automatically integrated out.
This approximation, and its derivatives, are obtained using automatic
differentiation (up to order three) of the joint likelihood.  The
computations are designed to be fast for problems with many random
effects ($\approx 10^6$) and parameters ($\approx
10^3$). Computation times using \pkg{ADMB} and \pkg{TMB} are compared
on a suite of examples ranging from simple models to large spatial
models where the random effects are a Gaussian random field.  Speedups
ranging from 1.5 to about 100 are obtained with increasing gains for
large problems.  The package and examples are available at
\url{http://tmb-project.org}.
}
\Plainkeywords{
	automatic differentiation,
	AD,
	random effects,
	latent variables,
  	C++ templates,
	R
}

\Address{
  Kasper Kristensen\\
  DTU Compute\\
  Matematiktorvet\\
  Building 303 B\\
  DK-2800 Kgs. Lyngby\\
  E-mail: \email{kaskr@imm.dtu.dk}\\
  URL: \url{http://imm.dtu.dk/~kaskr/}
}

\begin{document}
Submitted to \emph{Journal of Statistical Software}
\section{Introduction}
Calculation of derivatives plays an important role in computational
statistics.  One classic application is optimizing an objective
function; e.g., maximum likelihood.  Given a computer algorithm that
computes a function, Automatic Differentiation (AD)
\citep{griewank2008evaluating} is a technique that computes
derivatives of the function.  This frees the statistician from the
task of writing, testing, and maintaining derivative code.  This
technique is gradually finding its way into statistical software;
e.g., the \proglang{C++} packages \pkg{ADMB}
\citep{ADModelBuilder2011}, \pkg{Stan}~\citep{stan-software:2013} and
\pkg{Ceres Solver}~\citep{ceres-manual}.  These packages implement AD
from first principles, rather than using one of the general purpose AD
packages that are available for major programming languages such as
\proglang{Fortran}, \proglang{C++}, \proglang{Python}, and
\proglang{MATLAB} \citep{autodiff-tools}.  The Template Model Builder
(\pkg{TMB}) \proglang{R} package uses \pkg{CppAD} \citep{CppAD-manual}
to evaluate first, second, and possibly third order derivatives of a
user function written in \proglang{C++}.  Maximization of the
likelihood, or a Laplace approximation for the marginal likelihood, is
performed using conventional \proglang{R} optimization routines.  The
numerical linear algebra library \pkg{Eigen}~\citep{eigenweb} is used
for \proglang{C++} vector and matrix operations.  Seamless integration
of \pkg{CppAD} and \pkg{Eigen} is made possible through the use of
\proglang{C++} templates.

First order derivatives are usually sufficient for maximum likelihood
and for hybrid MCMC; e.g., the \pkg{Stan} package which provides both
only uses first order derivatives \citep{stan-software:2013}.  Higher
order derivatives calculated using AD greatly facilitate optimization
of the Laplace approximation for the marginal likelihood in complex
models with random effects; e.g., \citet{skaug_fournier1996aam}.  This
approach, implemented in \pkg{ADMB} and \pkg{TMB} packages, has been
used to fit simple random effect models as well as models containing
Gaussian Markov random fields (GMRF).  In this paper, we compare
computation times between these two packages for a range of random
effects models.

Many statisticians are unfamiliar with AD and for those we recommend
reading sections 2.1 and 2.2 of \citet{ADModelBuilder2011}.  The
\pkg{ADMB} package is rapidly gaining new users due to its superiority
with respect to optimization speed and robustness
\citep{bolker2013strategies} compared to e.g.~winBUGS
\citep{Spiegelhalter2003}.  The \pkg{TMB} package is built around the
same principles, but rather than being coded more or less from
scratch, it combines several existing high-performance libraries, to
be specific, \pkg{CppAD} for automatic differentiation in
\proglang{C++}, \pkg{Matrix} for sparse and dense matrix calculations
in \proglang{R}, \pkg{Eigen} for sparse and dense matrix calculations
in \proglang{C++}, and \pkg{OpenMP} for parallelization in
\proglang{C++} and \proglang{Fortran}.  Using these packages yields
better performance and a simpler code-base making \pkg{TMB} easy to
maintain.

The conditional independence structure in state-space models and GMRFs
yields a sparse precision matrix for the joint distribution of the
data and the random effects.  It is well known that, when this
precision matrix is sparse, it is possible to perform the Laplace
approximation for models with a very large number of random effects;
e.g., \pkg{INLA} \citep{rue2009approximate}.  The \pkg{INLA} package
(\url{http://www.r-inla.org/download}) restricts the models to cases
where the sparseness structure is known a~priori and models can be
written in one line of \proglang{R} code.  In contrast \pkg{ADMB}
requires manual identification of conditional independent likelihood
contributions, but is not restricted to any special model class.  The
\pkg{TMB} package can fit the same models as \pkg{ADMB}, but is better
equipped to take maximal advantage of sparseness structure.  It uses
an algorithm to automatically determine the sparsity structure.
Furthermore, in situations where the likelihood can be factored, it
enables parallelization using \pkg{OpenMP} \citep{dagum1998openmp}.
It also allows parallelization through \pkg{BLAS}
\citep{dongarra1990set} during Cholesky factorization of large sparse
precision matrices.  (Note that the \pkg{BLAS} library is written in
\proglang{Fortran}.)

\proglang{C++} templates treat variable types as parameters.  This
obtains the advantages of loose typing because one code base can work
on multiple types.  It also obtains the advantage of strong typing
because types are checked at compile time and the code is optimized
for the particular type.  \pkg{CppAD}'s use of templates enables
derivatives calculations to be recorded and define other functions
that can then be differentiated.  \pkg{Eigen}'s use of templates
enables matrix calculations where corresponding scalar types can do
automatic differentiation.  These features are important in the
implementation and use of \pkg{TMB}.

The rest of this paper is organized as follows:
Section~\ref{sec:Laplace} is a review of the Laplace approximation for
random effects models.  Section~\ref{sec:AD} is a review of automatic
differentiation as it applies to this paper.
Section~\ref{sec:Software} describes how \pkg{TMB} is implemented.
Section~\ref{sec:code_example} describes the package from a user's
perspective.  Section~\ref{sec:Case_studies} compares its performance
with that of \pkg{ADMB} for a range of models where the number of
parameters is between $1$ and $16$ and the number of random effects
is between $40$ and $40,000$.  Section~\ref{sec:Discussion} contains a
discussion and conclusion.

\section{The Laplace Approximation}
\label{sec:Laplace}
The statistical framework in this section closely follows that of
\citet{skaug_fournier1996aam}.  Let $f(u,\theta)$ denote the negative
joint log-likelihood of the data and the random effects.  This depends
on the unknown random effects $u \in \R^n$ and parameters $\theta
\in \R^m$.  The data, be it continuous or discrete, is not made
explicit here because it is a known constant for the analysis in this
section.  The function $f(u,\theta)$ is provided by the \pkg{TMB} user
in the form of \proglang{C++} source code.  The range of applications
is large, encompassing all random effects models for which the Laplace
approximation is appropriate.

The \pkg{TMB} package implements maximum likelihood estimation and
uncertainty calculations for $\theta$ and $u$.  It does this in an
efficient manner and with minimal effort on the part of the user.  The
maximum likelihood estimate for $\theta$ maximizes
\[  
  L(\theta)=\int_{\R^n}\exp(-f(u,\theta))\:du 
\]
w.r.t.~$\theta$.  Note that the random effects $u$ have been
integrated out and the marginal likelihood $L( \theta )$ is the
likelihood of the data as a function of just the parameters.  We
use $\hat{u} ( \theta )$ to denote the minimizer of $f(u, \theta)$
w.r.t.~$u$; i.e.,
\begin{equation}
  \hat u(\theta)=\arg\min_u f(u,\theta)
  \; .
  \label{def:u_hat}
\end{equation}
We use $H( \theta ) $ to denote the Hessian of $f ( u , \theta )$
w.r.t. $u$ and evaluated at $\hat{u} ( \theta )$; i.e.,
\begin{equation}
  H(\theta)=f''_{uu}(\hat u(\theta),\theta)
   \; .
  \label{def:H}
\end{equation}
The Laplace approximation for the marginal likelihood $L( \theta )$ is
\begin{equation}
  L^*(\theta) = \sqrt{2\pi}^n
   \det(H(\theta))^{-\frac{1}{2}}
     \exp(-f(\hat u,\theta)) 
  \; .
  \label{eq:laplace_approx}
\end{equation}
This approximation is widely applicable including models ranging from
non-linear mixed effects models to complex space-time models.  Certain
regularity conditions on the joint negative log-likelihood function
are required; e.g., the minimizer of $f(u , \theta )$ w.r.t.~$u$ is
unique.  These conditions are not discussed in this paper.

Models without random effects ($n=0$), and models for which the random
effects must be integrated out using classical numerical quadratures,
are outside the focus of this paper.

Our estimate of $\theta$ minimizes the negative log of the Laplace
approximation; i.e.,
\begin{equation}
  -\log L^*(\theta)=-n\log \sqrt{2\pi} 
                   + \frac{1}{2} \log \det(H(\theta)) + f(\hat u,\theta). 
  \label{eq:nll}
\end{equation}
This objective and its derivatives are required so that we can apply
standard nonlinear optimization algorithms (e.g., BFGS) to optimize
the objective and obtain our estimate for $\theta$.  Uncertainty of
the estimate $\hat\theta$, or any differentiable function of the
estimate $\phi(\hat\theta)$, is obtained by the $\delta$-method:
\begin{equation}
  \VAR(\phi(\hat \theta)) = -\phi'_\theta(\hat\theta)\left(\nabla^2
    \log L^*(\hat\theta)\right)^{-1}\phi'_\theta(\hat\theta)^\top.
  \label{eq:Vtheta}
\end{equation}
A generalized version of this formula is used to include cases where
$\phi$ also depends on the random effects, i.e.~$\phi(u,\theta)$,
\citep{skaug_fournier1996aam,kass1989approximate}. These uncertainty
calculations also require derivatives of \eqref{eq:nll}.  However,
derivatives are not straight-forward to obtain using automatic
differentiation in this context.  Firstly, because $\hat u$ depends on
$\theta$ indirectly as the solution of an \emph{inner} optimization
problem; see (\ref{def:u_hat}).  Secondly, equation~(\ref{eq:nll})
involves a log determinant, which is found through a Cholesky
decomposition.  A naive application of AD, that ignores sparsity,
would tape on the order of $n^3$ floating point operations.  While
some AD packages would not record the zero multiplies and adds, they
would still take time to detect these cases.  \pkg{TMB} handles these
challenges using state-of-the-art techniques and software packages.
In the next section, we review its use of the \pkg{CppAD} package for
automatic differentiation.

 \section{AD and CppAD}
\label{sec:AD}
Given a computer algorithm that defines a function, Automatic
Differentiation (AD) can be used to compute derivatives of the
function.  We only give a brief overview of~AD, and refer the reader
to \citet{griewank2008evaluating} for a more in-depth discussion.
There are two different approaches to AD: ``source transformation" and
``operator overloading".  In source transformation; e.g., the package
\pkg{TAPENADE} \citep{hascoet2004tapenade}, a preprocessor generates
derivative code that is compiled together with the original program.
This approach has the advantage that all the calculations are done in
compiler native floating point type (e.g., double-precision) which
tends to be faster than AD floating point types.  In addition the
compiler can apply its suite of optimization tricks to the derivative
code.  Hence source transformation tends to yield the best run time
performance, both in terms of speed and memory use.

In the operator overloading approach to AD, floating point operators
and elementary functions are overloaded using types that perform AD
techniques at run time.  This approach is easier to implement and to
use because it is not necessary to compile and interface to extra
automatically generated source code each time an algorithm changes.
\pkg{ADOL-C} \citep{Walther2012Gsw} and
\pkg{CppAD}~\citep{CppAD-manual} implement this approach using the
operator overloading features of \proglang{C++}.  Because \pkg{TMB}
uses \pkg{CppAD} it follows that its derivative calculations are based
on the operator overloading approach.

\begin{figure}[h!]
  \centering
  \includegraphics{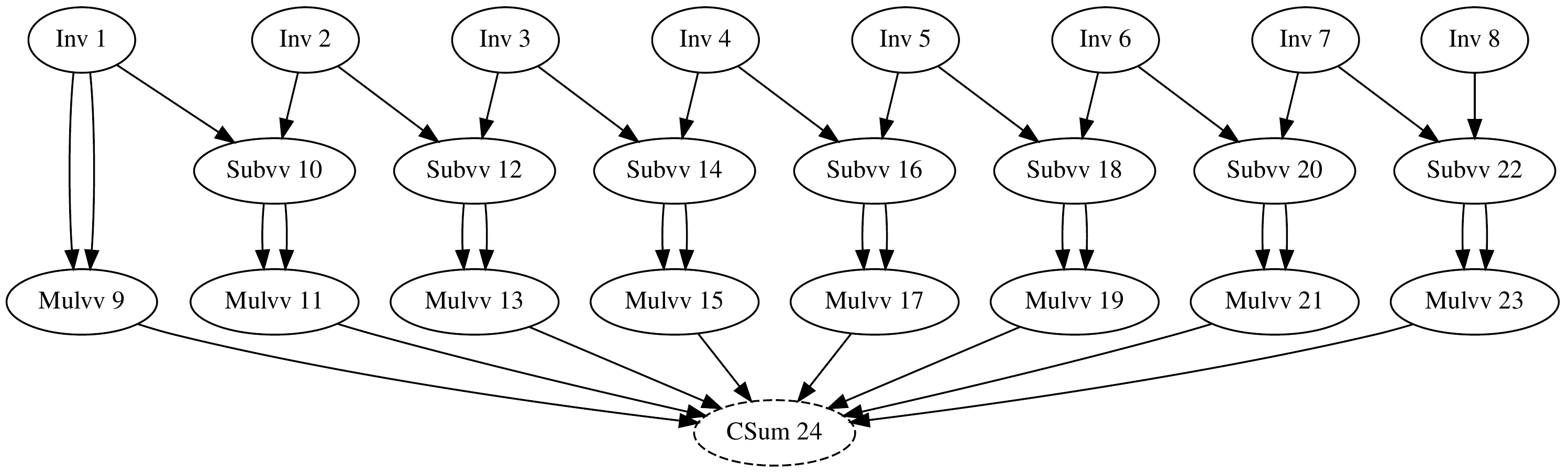} \\
  \caption{
    \pkg{CppAD} tape T1 for 
    $f ( \xi_1 , \ldots , \xi_8 ) 
    = 
    \xi_1^2 + \sum_{i=2}^8 ( \xi_i - \xi_{i-1} )^2
    \; .
    $ 
    Nodes ``Inv 1"--``Inv 8" correspond to $\xi_1 , \ldots , \xi_8$ and
    node ``CSum 24" corresponds to $f ( \xi_1 , \ldots , \xi_8 )$.
    Node labels indicate the elementary operations,
    numbering indicates the order in these operations are evaluated,
    arrows point from operation arguments to results,
    double arrows correspond to the square operator \code{x\^{}2} 
    which is implemented as \code{x*x}.
  }
  \label{fig:tapea}
\end{figure}

During evaluation of a user's algorithm, \pkg{CppAD} builds a
representation of the corresponding function, often referred to as a
``tape" or the ``computational graph".  Figure~\ref{fig:tapea} shows a
graphical representation of T1, the tape for the example function $f :
\R^8 \rightarrow \R$, defined by
\[
f( \xi_1,\ldots, \xi_8) =  \xi_1^2 + \sum_{i=2}^8( \xi_{i}- \xi_{i-1})^2 \; . 
\]
Each node corresponds to a variable, its name is the operation that
computes its value, and its number identifies it in the list of all
the variables.  The initial nodes are the independent variables $\xi_1
, \ldots \xi_8$.  The final node is the dependent variable
corresponding to the function value.  There are two main AD algorithms
known as the ``forward" and ``reverse" modes.  Forward mode starts
with the independent variables and calculates values in the direction
of the arrows.  Reverse mode does its calculations in the opposite
direction.

Because $f$ is a scalar valued function, we can calculate its
derivatives with one forward and one reverse pass through the
computational graph in Figure~\ref{fig:tapea}: Starting with the value
for the independent variables nodes 1 through 8, the forward pass
calculates the function value for all the other nodes.

The reverse pass, loops through the nodes in the opposite direction.
It recursively updates the $k$'th node's partial derivative $\partial
\xi_{24}/\partial \xi_k$, given the partials of higher nodes $\partial
\xi_{24}/\partial \xi_i$, for $i = k+1,...,24$.  For instance, to
update the partial derivative of node $k=5$, the chain rule is applied
along the outgoing edges of node 5; i.e.,
\[\frac {\partial \xi_{24}}{\partial \xi_{5}} = 
\frac {\partial \xi_{24}}{\partial \xi_{16}} 
\frac {\partial \xi_{16}}{\partial \xi_{5}}
+
\frac {\partial \xi_{24}}{\partial \xi_{18}} 
\frac {\partial \xi_{18}}{\partial \xi_{5}}
\] 
The partials of the final node, $\partial \xi_{24}/\partial \xi_{16}$
and $\partial \xi_{24}/\partial \xi_{18}$, are available from previous
calculations because 16 and 18 are greater than 5.  The partials along
the outgoing arrows, $\partial \xi_{16}/\partial \xi_{5}$ and
$\partial \xi_{18}/\partial \xi_{5}$, are derivatives of elementary
operations.  In this case, the elementary operation is subtraction and
these partials are plus and minus one.  (For some elementary
operations; e.g., multiplication, the values computed by the forward
sweep are needed to compute the partials of the elementary operation.)
On completion of the reverse mode loop, the total derivative of
$\xi_{24}$ w.r.t. the independent variables is available as $\partial
\xi_{24}/\partial \xi_{1}$,..., $\partial\xi_{24}/\partial \xi_{8}$.

For a scalar valued $f$, evaluation of its derivative using reverse
mode is surprisingly inexpensive.  The number of floating point
operations is less than~4 times that for evaluating $f$ itself
\citep{griewank2008evaluating}.  We refer to this as the ``cheap
gradient principle''.  This cost is proportional to the number of
nodes in T1; i.e., Figure~\ref{fig:tapea}.  (The actual result is for
a computational graph where there is only one or two arrows into each
node.  In T1, \pkg{CppAD} combined multiple additions into the final
node number 24.)  This result does not carry over from scalar valued
functions to vector valued functions $g(x)$.  It does apply to the
scalar-valued inner product $w^\top g(x)$, where $w$ is a vector in
the range space for $g$.  \pkg{CppAD} has provision for using reverse
mode to calculate the derivative of $w^\top g(x)$ given a range space
direction $w$ and a tape for $g(x)$.

\begin{figure}[h!]
  \centering
  \includegraphics{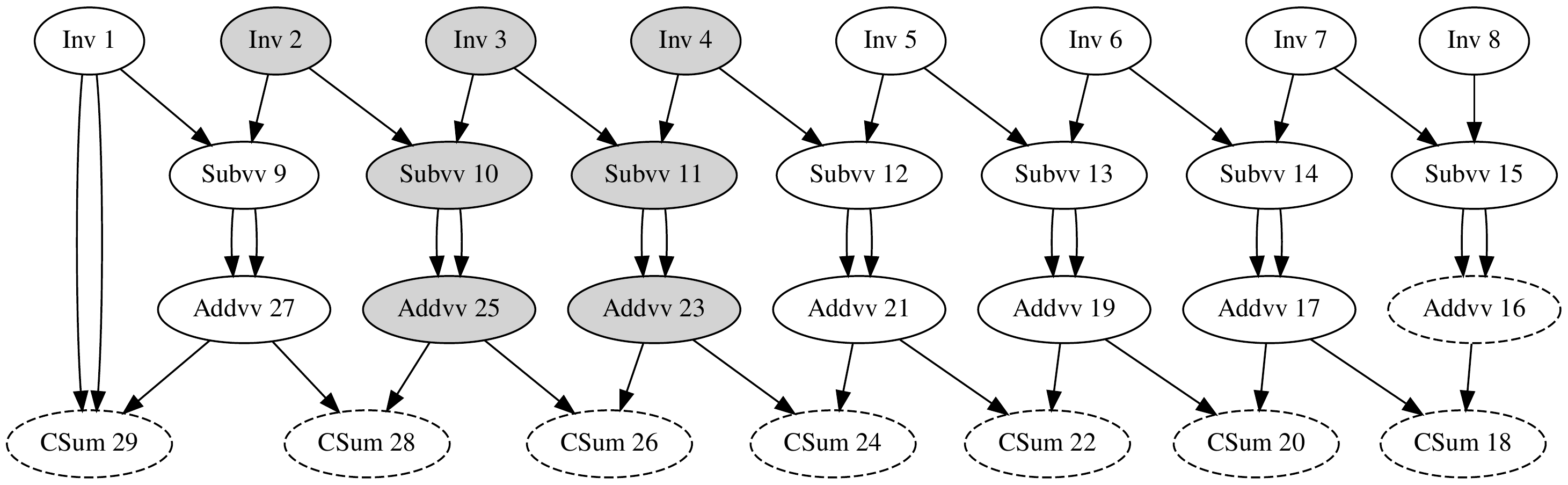} \\
  \caption{
    \pkg{CppAD} tape T2 for $f' ( \xi )$, when $f(\xi)$ is defined as in Figure~\protect\ref{fig:tapea}.
    For example, node~26 corresponds to the partial of $f$ w.r.t.~$\xi_3$; i.e.,
    $ f'_3 ( \xi ) = 2 ( \xi_3 - \xi_2 ) - 2 ( \xi_4 - \xi_3 )$.
    After a zero order forward sweep, $f'(\xi)$ is
    contained in nodes marked with dashed ellipses.
    Hessian columns of $f$ are found using first order reverse sweeps of T2. 
    For example, to find the 3rd Hessian column it is
    sufficient to traverse the sub-graph marked in gray; i.e., the nodes that
    affect the value of node 26.
  }
  \label{fig:tapeb}
\end{figure}

\pkg{CppAD} was chosen for AD calculations in \pkg{TMB} because it
provides two mechanisms for calculating higher order derivatives.  One
uses forward and reverse mode of any order.  The other is its ability
to tape functions that are defined in terms of derivatives and then
apply forward and reverse mode to compute derivatives of these
functions.  We were able to try many different derivative schemes and
choose the one that was fastest in our context.  To this end, it is
useful to tape the reverse mode calculation of $f'$ and thereby create
the tape T2 in Figure~\ref{fig:tapeb}.  This provides two different
ways to evaluate $f'$.  The new alternative is to apply a zero order
forward sweep on T2; i.e., starting with values for nodes 1-8,
sequentially evaluate nodes 9-29.  On completion the eight components
of the vector $f'$ are found in the dashed nodes of the graph.  If we
do a reverse sweep on~T2 in the direction $w$, we get
\begin{equation}
\partial_\xi [ w^\top f' ( \xi ) ]
=
\left(
\sum_{i=1}^8 w_i \frac{\partial^2}{\partial \xi_1 \partial \xi_i}  f(x)
\; , \; \ldots \; , \;
\sum_{i=1}^8 w_i \frac{\partial^2}{\partial \xi_8 \partial \xi_i}  f(x)
\right)
\label{eq:wT2}
\end{equation}
In Section~\ref{sec:Software} we shall calculate up to third order
derivatives using these techniques.

\newcommand{\mylabel}[1]{#1}

\pkg{CppAD} does some of its optimization during the taping procedure;
e.g., constant expressions are reduced to a single value before being
taped.  Other optimizations; e.g., removing code that does not affect
the dependent variables, can be performed using an option to optimize
the tape.  This brings the performance of \pkg{CppAD} closer to the
source transformation AD tools, especially in cases where the
optimized tape is evaluated a large number of times (as is the case
with \pkg{TMB}).

\section{Software Implementation}
\label{sec:Software}
\pkg{TMB} calculates estimates of both parameters and random effects
using the Laplace approximation~(\ref{eq:laplace_approx}) for the
likelihood.  The user provides a \proglang{C++} function $f(u,
\theta)$ that computes the joint likelihood as a function of the
parameters $\theta$ and the random effects $u$; see
Section~\ref{sec:code_example} for more details.  This function,
referred to as the ``user template" below, defines the user's
statistical model using a standard structure that is expected by the
package.  Its floating point type is a template parameter so that it
can be used with multiple \pkg{CppAD} types.  Hence there are two
meanings of the use of T (Template) in the package name \pkg{TMB}.

\begin{figure}
  \centering
\includegraphics{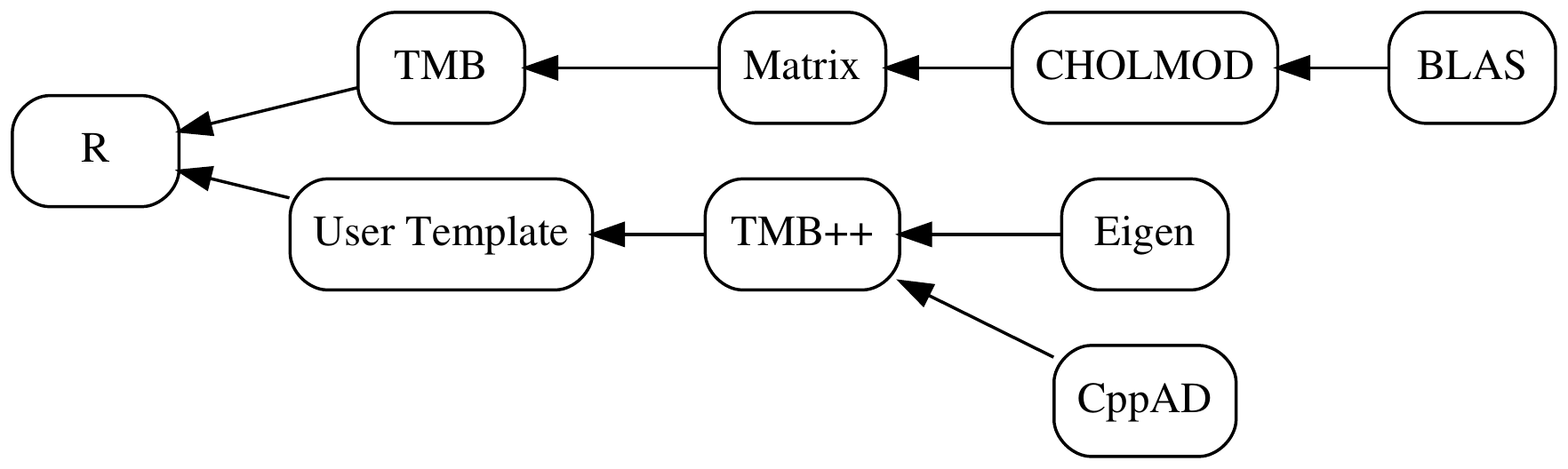}
\caption{\pkg{TMB} package design:
module R is the the top level controlling user code,
TMB is the part of \pkg{TMB} that is written in \proglang{R},
\pkg{Matrix} is an \proglang{R} sparse matrix package,
\pkg{CHOLMOD} is a \proglang{C} sparse Cholesky factorization routine,
\pkg{BLAS} is a \proglang{Fortran} tuned set of basic linear algebra routines,
User Template is the user's joint likelihood in \proglang{C++},
TMB++ is the package components in \proglang{C++},
\pkg{Eigen} is a \proglang{C++} matrix algebra library,
\pkg{CppAD} is a \proglang{C++} AD package. Arrows indicate package inclusions.
}
\label{fig:package_graph}
\end{figure}

An overview of the package design is shown in Figure
\ref{fig:package_graph}.  Evaluation of the objective $-\log
L^*(\theta)$ and its derivatives, is performed in \proglang{R}; see
equation~(\ref{eq:nll}).  \pkg{TMB} performs the Laplace approximation
with use of \pkg{CHOLMOD}, natively available in \proglang{R} through
the \pkg{Matrix} package, and optionally linking to \pkg{BLAS}.
Sub-expressions, such as $\hat{u} ( \theta )$ and $H ( \theta )$, are
evaluated in \proglang{C++}.  These sub-expressions are returned as
\proglang{R} objects, and the interactive nature of \proglang{R}
allows the user to easily inspect them.  This is important during a
model development stage.

Interfaces to the various parts of \pkg{CppAD} constitute a large part
of the \proglang{R} code. During an initial phase of program execution
the following \pkg{CppAD} tapes are created:
\begin{itemize}
\item[T1]
  Tape of $f(u,\theta) : \R^{m+n}\rightarrow \R$, 
   generated from user program. 
  Graph size proportional to flop count of user template function.
\item[T2]
  Tape of $f' (u,\theta) : \R^{m+n}\rightarrow \R^{m+n}$,
  generated from T1 as described in Figure~\ref{fig:tapeb}. 
  Graph size is at most 4 times the size of T1.
 \item[T3] Tape of $f''_{uu} (u,\theta) : \R^{m+n} \rightarrow \R^\ell$, 
  the $l$~non-zero entries in the lower triangle of $H( \theta )$.
  Prior to T3's construction,
  the sparsity pattern of $H( \theta )$ is calculated by analyzing 
  the dependency structure of T2.
\end{itemize}
Tapes T1-T3 correspond to Codes 1-3 in Table~1
of~\citet{skaug_fournier1996aam}.  These tapes are computed only once
and are subsequently held in memory.  The corresponding data
structures are part of the \proglang{R} environment and are managed by
the \proglang{R} garbage collector just like any other objects created
from the \proglang{R} command line.

\subsection{Inverse Subset Algorithm}
\label{sec:inverse_subset}
In this section we describe how the tapes T1-T3 are used to calculate
the derivative of the objective with respect to the parameters.
Define $h : \R^{n+m} \rightarrow \R$ by
\[
h(u,\theta) =
-\frac{n}{2}\log{2\pi}+\frac{1}{2}\log{\det{f''_{uu}(u,\theta)}}+f(u,\theta)
\; .
\]
It follows that the objective $-\log L^*(\theta)$ is equal to $h(\hat
u(\theta),\theta)$.  Furthermore, the function $\hat{u} ( \theta )$
satisfies the equations
\begin{eqnarray*}
f'_u ( \hat{u} ( \theta ) , \theta ) & = & 0
\\
\hat{u}' ( \theta ) & = & 
- 
f''_{uu} ( \hat{u} ( \theta ) , \theta )^{-1}
f''_{u \theta } ( \hat{u} ( \theta ) , \theta ).
\end{eqnarray*}
The derivative of the objective w.r.t. $\theta$ is 
\begin{equation}
  \label{gradient}
\frac{d}{d\theta} h( \hat{u} ( \theta) , \theta )
=
h'_{\theta}( \hat{u} ( \theta ) , \theta )
-
h'_{u}( \hat{u} ( \theta ) , \theta )
f''_{uu} ( \hat{u} ( \theta ) , \theta )^{-1}
f''_{u \theta } ( \hat{u} ( \theta ) , \theta ).
\end{equation}

To simplify the notation, we express $( u , \theta)$ as a single
vector $\xi \in \R^{n+m}$.  The first step is to evaluate
\[
h'_{\xi} (\xi)
= 
f'_{\xi}(\xi)
+
\frac{d}{d \xi} \frac{1}{2} \log{\det{f''_{uu}(\xi)}}.
\]
The term $f'_\xi ( \xi )$ is calculated using a first order reverse
sweep on T1.  The derivative of the log-determinant is calculated
using the well known rule
\begin{equation}
  \label{eq:partial_of_logdet}
\frac{\partial}{\partial \xi_i}\left(
  \frac{1}{2}\log{\det{f''_{uu}(\xi)}} \right)=
\frac{1}{2}{\rm tr} \left( f''_{uu}(\xi)^{-1}  \frac{\partial}{\partial
    \xi_i} f''_{uu}(\xi) \right).
\end{equation}
The trace of a product of symmetric matrices ${\rm tr} (A B)$ is equal
to the sum of the entries of the pointwise product $A\odot B$.  Thus,
computing the right hand side of equation~(\ref{eq:partial_of_logdet})
only requires the elements of $f''_{uu}(\xi)^{-1}$ that correspond to
non-zero entries in the sparsity pattern for $\partial_{\xi(i)}
f''_{uu} ( \xi )$.  The \emph{inverse subset algorithm} transforms the
sparse Cholesky factor $L$ to the inverse $(LL^\top)^{-1}$ on the
sparseness pattern of $LL^\top$; e.g., \citet{rue2005marginal}.  Let
$w \in \R^\ell$ denote the elements of $f''_{uu} ( \xi )^{-1}$ that
correspond to non-zeros in the lower triangle of $f''_{uu}(\xi)$.  We
can compute the partial~(\ref{eq:partial_of_logdet}), for all $i$,
through a single first order reverse sweep of tape T3 in range
direction $w$.

Having evaluated $h'(\xi)$ we turn to the remaining terms in
equation~(\ref{gradient}).  A sparse matrix-vector solve is used to
compute
\[
v = 
h'_{u}( \hat{u} ( \theta ) , \theta )
f''_{uu} ( \hat{u} ( \theta ) , \theta )^{-1} \; .
\]
A reverse mode sweep of tape T2 in range direction $w=(v,0)$ is used
to compute
\[
\frac{d}{d\xi} [ v f'_u ( \hat{u} ( \theta ) , \theta ) ]
= 
[ 
v f''_{uu} ( \hat{u} ( \theta ) , \theta ) , 
v f''_{u \theta } ( \hat{u} ( \theta ) , \theta ) 
]
\; .
\]
This yield the final term needed in equation~(\ref{gradient})
\[
h'_{u}( \hat{u} ( \theta ) , \theta )
f''_{uu} ( \hat{u} ( \theta ) , \theta )^{-1} 
f''_{u \theta } ( \hat{u} ( \theta ) , \theta ) 
=
v f''_{u \theta } ( \hat{u} ( \theta ) , \theta ) 
\; .
\]
Note that the term $ v f''_{uu} ( \hat{u} ( \theta ) , \theta ) $ is
not used by the method above.  It is necessary to include $u$ in the
independent variables for this calculation so that the dependence of $
v f''_{u \theta } ( \hat{u} ( \theta ) , \theta ) $ on the value of
$\hat{u} ( \theta )$ can be included.

\newcommand{\vempty}[1]{\raisebox{0em}[#1][0em]{} }
\begin{table}[ht]
\centering
\begin{tabular}{|lc|lc|}
\hline
Laplace:                 & \vempty{1.1em} $h( \hat{\xi} )$
&  
Gradient:                & $ ( d / d \theta ) h ( \hat{\xi} ) $
\\
\hline
L1: min $f( u , \theta )$ w.r.t. $u$ & $\hat{\xi} = ( \hat{u} , \theta)$
& &  
\\
L2: order 0 forward T1   &  $ f( \hat{\xi} ) $ 
& 
G1: order 1 reverse T1 & $ f' ( \hat{\xi} ) $
\\
L3: order 0 forward T3   & $ f''_{uu} ( \hat{\xi} ) $
&  
G2: order 1 reverse T3 & $ \partial_\xi f''_{uu} ( \hat{\xi} ) $
\\
L4: sparse Cholesky      & $ L L^\top = f''_{uu} ( \hat{\xi} ) $  &  
G3: inverse subset       & $\partial_\xi \log \det f''_{uu} ( \hat{\xi} )$
\\
&&  
G4: sparse solve         & $ v = h'_u ( \hat{\xi} ) f_{uu} ( \hat{\xi} )^{-1} $
\\
&&  

G5: order 0 forward T2   & $ f' ( \hat{\xi} ) $
\\
&&  
G6: order 1 reverse T2 & $ v f''_{u , \theta} ( \hat{\xi} ) $
\\
\hline
\end{tabular}
\caption{Computational steps for the Laplace approximation and its gradient in \pkg{TMB}.}
\label{tab:comp_steps}
\end{table}
\newcommand{\work}[1]{\text{work}\left(\text{#1}\right)} 
The computational steps for evaluating the Laplace approximation and its
gradient are summarized in Table \ref{tab:comp_steps}.  Note that the
G1 calculation of $f' ( \hat{\xi} )$ could in principle be avoided by
reusing the result of G5. However, as the following work calculation
shows, the overall computational approach is efficient.  The work
required to evaluate the Laplace approximation is
\[
\work{Laplace} = \work{L1+L2+L3+L4},
\]
while the the work of the entire table is
\[
\work{Laplace+Gradient}
= 
\work{L1+L2+L3+L4+G1+G2+G3+G4+G5+G6}.
\]
It follows from the cheap gradient principle that,
\[
\work{Laplace+Gradient}
\leq
\work{L1+L4+G3+G4}
+
4 \cdot \work{L2+L3+G5}.
\]
Given the sparse Cholesky factorization, the additional work required
for the inverse subset algorithm is equal to the work of the sparse
Cholesky factorization \citep{YoginAndDavis:TR-95-021}.  The
additional work required for the sparse solve is less than or equal
the work for the sparse Cholesky factorization.  We conclude that
\begin{eqnarray*}
\work{L4+G3+G4} & \leq & 3 \cdot \work{L4},
\\
\work{Laplace+Gradient}
& \leq &
\work{L1}
+
4 \cdot \work{L2+L3+L4+G5}.
\end{eqnarray*}
Under the mild assumption that solving the inner problem, L1, requires
at least two evaluations of $f'( \xi )$, i.e., $2\cdot\work{G5}\leq
\work{L1}$, we conclude
\[
\work{Laplace+Gradient} 
\leq 
4 \cdot \work{L1+L2+L3+L4} 
=
4 \cdot \work{Laplace}.
\]
Hence, the cheap gradient principle is preserved for the gradient of
Laplace approximation.

Besides from efficient gradient calculations, the inverse subset
algorithm is used by \pkg{TMB} to calculate marginal standard
deviations of random effects and parameters using the generalized delta
method \citep{kass1989approximate}, which is also used in \pkg{ADMB}.

\subsection{Automatic Sparsity Detection}
\label{sec:sparseness_detection}
\pkg{TMB} can operate on very high dimensional problems provided that
the Hessian~(\ref{def:H}) is a sparse matrix.  In this section, we
illustrate how the sparsity structure of $H$ is automatically detected
and comment on the computational cost of this detection.

Consider the negative joint log-likelihood for a one dimensional
random walk with $\M{N}(0,\frac{1}{2})$ innovations and no
measurements:
\[
f( u , \theta )
=
u_0^2 + \sum_{i=1}^7 ( u_i - u_{i-1} )^2
\; .
\]
The bandwidth of the Hessian $f''_{uu} ( u , \theta )$ is three.
Below is a user template implementation of this negative joint
log-likelihood.  (Refer to Section~\ref{sec:code_example} for details
about the structure of a \pkg{TMB} user template.)
\begin{Code}
#include <TMB.hpp>
template<class Type>
Type objective_function<Type>::operator() ()
{ 
  PARAMETER_VECTOR(u);
  Type f = pow(u[0], 2);
  for(int i = 1; i < u.size(); i++) f += pow(u[i] - u[i-1], 2);
  return f;
}
\end{Code}
Figure~\ref{fig:tapea} shows a tape T1 corresponding to this user
template.  The nodes are numbered (1 to 24) in the order they are
processed during a forward sweep.  \pkg{CppAD} is used to record the
operations that start with the independent variables nodes 1 to 8 of
T1, perform a zero order forward sweep and then a first order reverse
sweep, and result in the derivative of $f$ (in nodes 1 to 8 of T1).
This recording is then processed by \pkg{CppAD}'s tape optimization
procedure and the result is tape T2 (Figure \ref{fig:tapeb}).  In this
tape, the input values are numbered 1 to 8 (as in T1) and the output
values are the dashed nodes in the last row together with node 16.

If we take $w$ in equation~(\ref{eq:wT2}) to be the $k$th unit vector,
a reverse sweep for~T2 will yield the $k$th column of the Hessian $H$
of $f$.  However, such full sweeps are far from optimal.  Instead, we
find the subgraph that affects the $k$th gradient component, and
perform the reverse sweep only on the subgraph.  The example $k=3$ is
shown in Fig~\ref{fig:tapeb} where the dependencies of node 26 (3rd
gradient component) are marked with gray.  \pkg{TMB} determines the
subgraph using a breadth-first search from the $k$th node followed by
a standard sort.  This gives a computational complexity of $O(n_k \log
n_k)$ where $n_k$ is the size of the $k$th subgraph.

A further reduction would be possible by noting that the sorting
operation can be avoided: The reverse sweep need not be performed in
the order of the original graph. A topological sort is sufficient, in
principle reducing the computational complexity to $O(n_k)$.  For a
general quadratic form the computational complexity can in theory
become as low as proportional to the number of non-zeros of the
Hessian. At worst, for a dense Hessian, this gives a complexity of
$O(n^2)$ (though the current implementation has $O(n^2 \log n)$). In
conclusion, the cost of the sparse Hessian algorithm is small compared
to e.g., the Cholesky factorization. Also recall that the Hessian
sparseness detection only needs to be performed once for a given
estimation problem.

\subsection{Parallel Cholesky through BLAS}
\label{sec:BLAS}
For models with a large number of random effects, the most demanding
part of the calculations is the sparse Cholesky factorization and the
inverse subset algorithm; e.g., when the random effects are
multi-dimensional Gaussian Markov Random Fields (see
Section~\ref{sec:code_example}).  For such models the work of the
Cholesky factorization is much larger than the work required to build
the Hessian matrix $f''_{uu}( \xi )$ and to perform the AD
calculations.  The \pkg{TMB} Cholesky factorization is performed by
\pkg{CHOLMOD} \citep{chen2008algorithm}, a supernodal method that uses
the \pkg{BLAS}.  Computational demanding models with large numbers of
random effects can be accelerated by using parallel and tuned
\pkg{BLAS} with the \proglang{R} installation; e.g., MKL
\citep{intel2007intel}.  The use of parallel \pkg{BLAS} does not
improve performance for models where the Cholesky factor is very
sparse (e.g., small-bandwidth banded Hessians), because the \pkg{BLAS}
operations are then performed on scalars, or low dimensional dense
matrices.

\subsection{Parallel User Templates using OpenMP}
\label{sec:parallel}
\begin{figure}[htbp]
  \centering
  $\begin{array}{c}
  \mylabel{(a)} \\ 
  \includegraphics{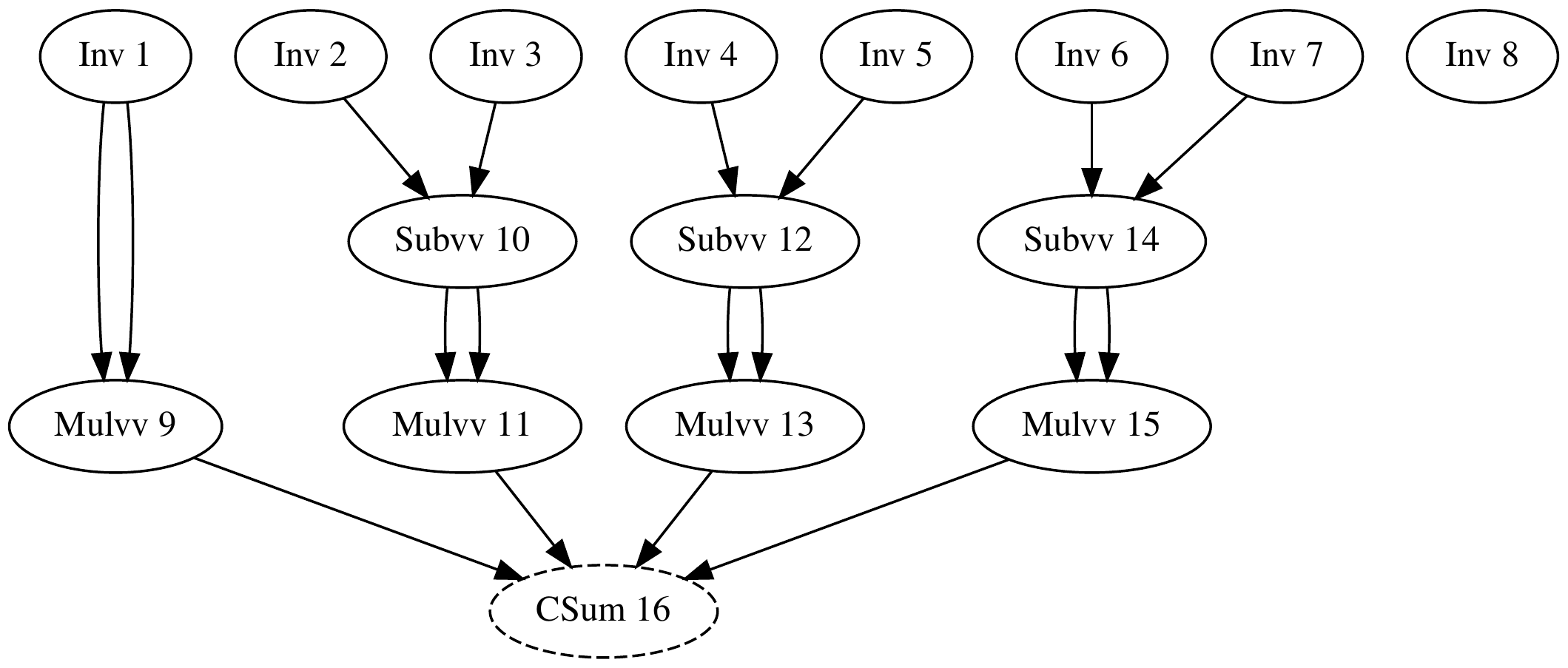} \\
  \mylabel{(b)} \\ 
  \includegraphics{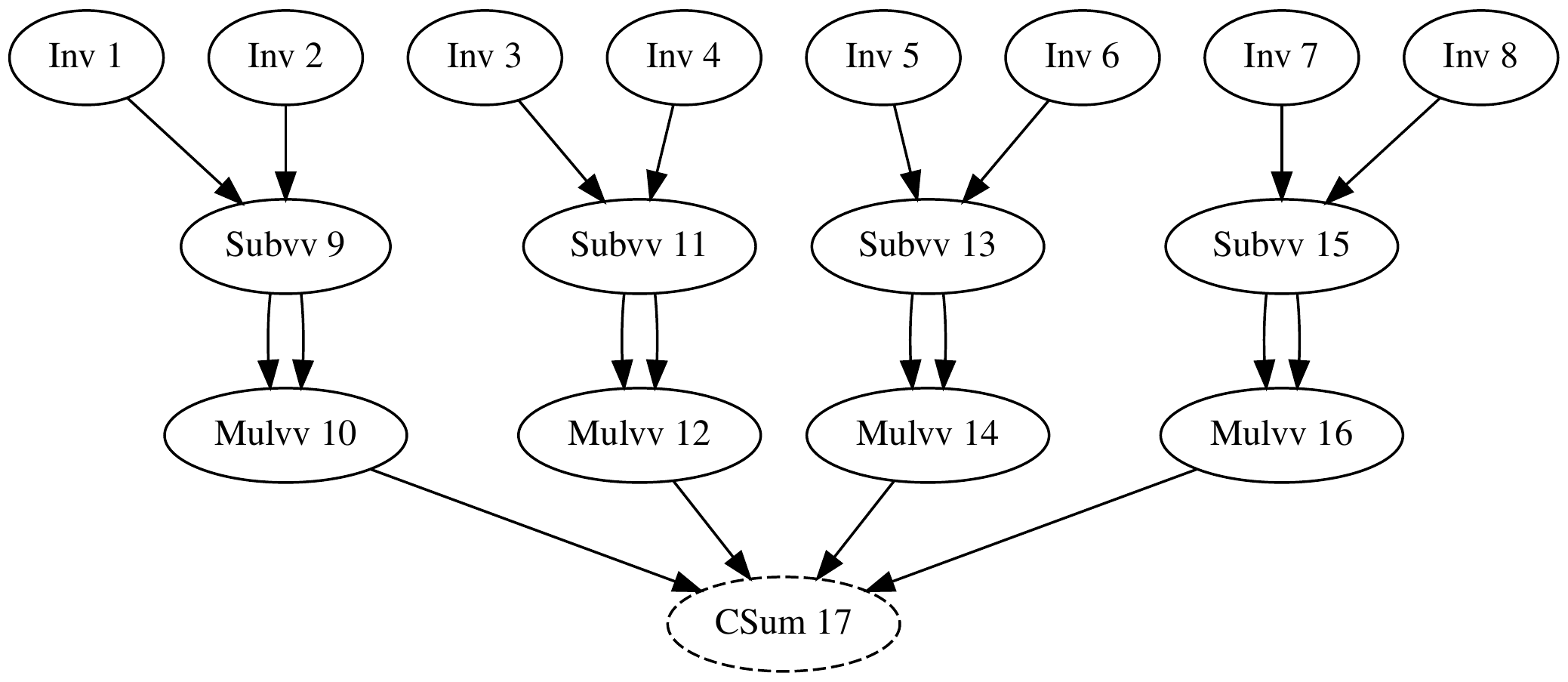} \\
  \end{array}$
  \caption{Illustration of automatic parallelization. After change of
    return type to \code{parallel\_accumulator} the tape of Figure \ref{fig:tapea} is split such
    that thread 1 accumulates the ``even'' terms and thereby generates
    the tape (a) and thread 2 accumulates the ``odd'' terms thereby generating the
    tape (b). The sum of node 16 (a) and node 17 (b) gives the same result as
    node 24 of Figure \ref{fig:tapea}. All further AD are processed
    independently by the threads including sparsity detection and Hessian
    calculations. \pkg{TMB} glues the results together from the individual threads.}
\label{fig:tape_parallel}
\end{figure}

For some models the evaluation of $f(\xi)$ is the most time consuming
part of the calculations.  If the joint likelihood corresponds to
independent random variables, $f( \xi )$ is a result of summation;
i.e.,
\[
	f( \xi ) = \sum_{k=1}^K f_k (\xi).
\]
For example, in the case of a state-space model, $f_k ( \xi )$ could
be the negative log-likelihood contribution of a state transition for
the $k$th time-step.  Assume for simplicity that two computational
cores are available.  We could split the sum into even and odd values
of $k$; i.e.,
\[
	f(\xi) = f_\text{even} (\xi) + f_\text{odd} (\xi) \; .
\]
We could use any other split such that the work of the two terms are
approximately equal.  All AD calculations can be performed on
$f_\text{even}(\xi)$ and $f_\text{odd}(\xi)$ separately in parallel
using \pkg{OpenMP}.  This includes construction of the tapes T1, T2,
T3, sparseness detection, and subsequent evaluation of these tapes.
The parallelization targets all computational steps of Table
\ref{tab:comp_steps} except L4, G3 and G4.  As an example, consider
the tape T1 in Figure~\ref{fig:tapea}; i.e., tape T1 for the simple
random walk example.  In the two core case this tape would be split as
shown in Figure~\ref{fig:tape_parallel} (Appendix).  In general for
any number of cores, if the user template includes
\code{parallel\_accumulator<Type> f(this);}, \pkg{TMB} automatically
splits the summation using \code{f +=} and \code{f -=} and computes
the sum components in parallel; see examples in the results section.

\section{Using the TMB Package}
\label{sec:code_example}
Using the \pkg{TMB} package involves two steps that correspond to the
User Template and R boxes in Figure~\ref{fig:package_graph}.  The User
Template defines the negative joint log-likelihood using specialized
macros that pass the parameters, random effects, and data from
\proglang{R}.  The R box typically prepares data and initial values,
links the user template, invokes the optimization, and post processes
the results returned by the TMB box.  The example below illustrates
this process.

Consider the ``theta logistic'' population model of
\citep{wang2007example} and \citep{pedersen2011estimation}. This is a
state-space model with state equation
\[
u_t 
=
u_{t-1} + r_0\left(1-\left(\frac{\exp(u_{t-1})}{K}\right)^\psi\right)
+
e_t,
\]
and observation equation
\[
y_t = u_t + v_t,
\]
where $e_t \sim \M{N}(0,Q)$, $v_t \sim \M{N}(0,R)$ and $t\in
\{0,...,n-1\}$.  All of the state values $u_0 , \ldots , u_{n-1}$ are
random effects and integrated out of the likelihood.  A uniform prior
is implicitly assigned to $u_0$.  The parameter vector is
$\theta=(\log(r_0),\log(\psi),\log(K),\log(Q),\log(R))$. The joint
density for $y$ and $u$ is
\[
\left(\prod_{t=1}^{n-1} p_\theta(u_t|u_{t-1})\right)
\left(\prod_{t=0}^{n-1} p_\theta(y_t|y_t)\right) \; .
\]
The negative joint log-likelihood is given by
\begin{eqnarray*}
f(u,\theta)
& = & 
- \sum_{t=1}^{n-1} \log p_\theta(u_t|u_{t-1})
- \sum_{t=0}^{n-1} \log p_\theta(y_t|u_t)
\\
& = &
- \sum_{t=1}^{n-1} \log p_\theta(e_t)
- \sum_{t=0}^{n-1} \log p_\theta(v_t).
\end{eqnarray*}
The user template for this negative joint log-likelihood (the file
named~\code{thetalog.cpp} at
\url{https://github.com/kaskr/adcomp/tree/master/tmb_examples}) is
\begin{Code}
#include <TMB.hpp>
template<class Type>
Type objective_function<Type>::operator() ()
{
  DATA_VECTOR(y);               // data
  PARAMETER_VECTOR(u);          // random effects
  // parameters
  PARAMETER(logr0);    Type r0 = exp(logr0);
  PARAMETER(logpsi);   Type psi = exp(logpsi);
  PARAMETER(logK);     Type K = exp(logK);
  PARAMETER(logQ);     Type Q = exp(logQ);
  PARAMETER(logR);     Type R = exp(logR);
  int n = y.size();             // number of time points
  Type f = 0;                   // initialize summation
  for(int t = 1; t < n; t++){   // start at t = 1
    Type mean = u[t-1] + r0 * (1.0 - pow(exp(u[t-1]) / K, psi));
    f -= dnorm(u[t], mean, sqrt(Q), true);   // e_t
  }
  for(int t = 0; t < n; t++){   // start at t = 0
    f -= dnorm(y[t], u[t], sqrt(R), true);   // v_t
  }
  return f;
}
\end{Code}
There are a few important things to notice.  The first four lines, and
the last line, are standard and should be the same for most models.
The first line includes the \pkg{TMB} specific macros and functions,
including dependencies such as \pkg{CppAD} and \pkg{Eigen}.  The
following three lines are the syntax for starting a function template
where \code{Type} is a template parameter that the compiler replaces
by an AD type that is used for numerical computations.  The line
\code{DATA_VECTOR(y)} declares the vector \code{y} to be the same as
\code{data\$y} in the \proglang{R} session (included below).  The line
\code{PARAMETER_VECTOR(u)} declares the vector \code{u} to be the same
as \code{parameters\$u} in the \proglang{R} session.  The line
\code{PARAMETER(logr0)} declares the scalar \code{logr0} to be the
same as \code{parameters\$logr0} in the \proglang{R} session.  The
other scalar parameters are declared in a similar manner.  Note that
the user template does not distinguish between the parameters and
random effects and codes them both as parameters.  The density for a
normal distribution is provided by the function \code{dnorm}, which
simplifies the code.  Having specified the user template it can be
compiled, linked, evaluated, and optimized from within \proglang{R}:
\begin{Code}
y <- scan("thetalog.dat", skip = 3, quiet = TRUE)
library("TMB")
compile("thetalog.cpp")
dyn.load(dynlib("thetalog"))
data <- list(y = y)
parameters <- list(
  u = data$y * 0,
  logr0 = 0,
  logpsi = 0,
  logK = 6,
  logQ = 0,
  logR = 0
  )
obj <- MakeADFun(data, parameters, random = "u", DLL = "thetalog")
system.time(opt <- nlminb(obj$par, obj$fn, obj$gr))
rep <- sdreport(obj)
\end{Code}
The first line uses the standard \proglang{R} function \code{scan} to
read the data vector \code{y} from a file.  The second line loads the
\pkg{TMB} package.  The next two lines compile and link the user
template.  The line \code{data <- list(y = y)} creates a data list for
passing to \code{MakeADFun}.  The data components in this list must
have the same names as the \code{DATA_VECTOR} names in the user
template.  Similarly a parameter list is created where the components
have the same names as the parameter objects in the user template.
The values assigned to the components of \code{parameter} are used as
initial values during optimization.  The line that begins \code{obj <-
  MakeADFun} defines the object \code{obj} containing the data,
parameters and methods that access the objective function and its
derivatives.  If any of the parameter components are random effects,
they are assigned to the \code{random} argument to \code{MakeADFun}.
For example, if we had used \code{random = c("u", "logr0")},
\code{logr0} would have also been a random effect (and integrated out
using the Laplace approximation.)  The last three lines use the
standard \proglang{R} optimizer \code{nlminb} to minimize the Laplace
approximation \code{obj\$fn} aided by the gradient \code{obj\$gr} and
starting at the point \code{obj\$par}.  The last line generates a
standard output report.
\section{Case Studies}
\label{sec:Case_studies}
A number of case studies are used to compare run times and accuracy
between \pkg{TMB} and \pkg{ADMB}; see Table~\ref{tab:description}.
These studies span various distribution families, sparseness
structures, and inner problem complexities.  Convex inner
problems~(\ref{def:u_hat}) are efficiently handled using a Newton
optimizer, while the non-convex problems generally require more
iterations and specially adapted optimizers.
\begin{table}[!htb]
\centering
\begin{tabular}{lp{5.5cm}rrp{1.5cm}l}
  \textbf{Name} & \textbf{Description} & \textbf{dim $u$} & \textbf{dim $\theta$} & \textbf{Hessian} & \textbf{Convex} \\
  \hline
  mvrw & Random walk with multivariate correlated increments and measurement noise & 300 & 7 & block(3) tridiagonal & yes \\
  nmix &  Binomial-Poisson mixture model \citep{royle2004n} & 40 & 4 & block(2) diagonal & no \\
  orange\_big & Orange tree growth example (\citet{pinheiro2000mixed}, Ch.8.2) & 5000 & 5 & diagonal & yes \\
  sam & State-space fish stock assessment model \citep{nielsen2014estimation} & 650 & 16 & banded(33) & no \\
  sdv\_multi & Multivariate SV model \citep{skaug2013flexible} & 2835 & 12 & banded(7) & no \\
  socatt &  Ordinal response model with random effects & 264 & 10 & diagonal & no \\
  spatial & Spatial Poisson GLMM on a grid, with exponentially decaying correlation function & 100 & 4 & dense & yes \\
  thetalog & Theta logistic population model \citep{pedersen2011estimation} & 200 & 5 & banded(3) & yes \\
  ar1\_4D &  Separable GMRF on 4D lattice with AR1 structure in each direction and Poisson measurements & 4096 & 1 & 4D Kronecker & yes \\
  ar1xar1 & Separable covariance on 2D lattice with AR1 structure in each direction and Poisson measurements & 40000 & 2 & 2D Kronecker & yes \\
  longlinreg & Linear regression with $10^6$ observations & 0 & 3 & - & - \\
\end{tabular}
\caption{
  Description of case studies and problem type, specifically
  number of random effects (dim $u$) and parameters (dim $\theta$), sparseness structure
  of the Hessian (\protect\ref{def:H}) and inner optimization problem
  type (convex/not convex). Source code for the examples are available
  at \url{https://github.com/kaskr/adcomp/tree/master/tmb_examples}.
}
\label{tab:description}
\end{table}

\subsection{Results}
\label{sec:Results}
\begin{table}[ht]
\centering
\begin{tabular}{lccccc}
  \hline
  Example & $r(\hat\theta_1,\hat\theta_2)$ & $r(\sigma(\hat\theta_1),\sigma(\hat\theta_2))$ & $r(\hat u_1,\hat u_2)$ & $r(\sigma(\hat u_1),\sigma(\hat u_2))$  \\
  \hline
  longlinreg & 0.003$\times 10^{-4}$ & 0.000$\times 10^{-4}$ &  &  \\
  mvrw & 0.156$\times 10^{-4}$ & 0.077$\times 10^{-4}$ & 0.372$\times 10^{-4}$ & 0.089$\times 10^{-4}$ \\
  nmix & 0.097$\times 10^{-4}$ & 0.121$\times 10^{-4}$ & 0.222$\times 10^{-4}$ & 0.067$\times 10^{-4}$ \\
  orange\_big & 0.069$\times 10^{-4}$ & 0.042$\times 10^{-4}$ & 0.026$\times 10^{-4}$ & 1.260$\times 10^{-4}$ \\
  sam & 0.022$\times 10^{-4}$ & 0.167$\times 10^{-4}$ & 0.004$\times 10^{-4}$ & 0.019$\times 10^{-4}$ \\
  sdv\_multi & 0.144$\times 10^{-4}$ & 0.089$\times 10^{-4}$ & 0.208$\times 10^{-4}$ & 0.038$\times 10^{-4}$ \\
  socatt & 0.737$\times 10^{-4}$ & 0.092$\times 10^{-4}$ & 0.455$\times 10^{-4}$ & 1.150$\times 10^{-4}$ \\
  spatial & 0.010$\times 10^{-4}$ & 0.160$\times 10^{-4}$ & 0.003$\times 10^{-4}$ & 0.001$\times 10^{-4}$ \\
  thetalog & 0.001$\times 10^{-4}$ & 0.007$\times 10^{-4}$ & 0.000$\times 10^{-4}$ & 0.000$\times 10^{-4}$ \\

  \hline
\end{tabular}
\caption{Comparison of TMB estimates (subscript 1) versus ADMB
  (subscript 2): parameters $\hat\theta$, parameters standard
  deviation $\sigma(\hat\theta)$, random effects $\hat u$,
  random effects standard deviation $\sigma(\hat u)$, using the
  distance measure
  $r(x,y)=2\|x-y\|_{\infty}/(\|x\|_{\infty}+\|y\|_{\infty})$.
}
\label{tab:compare_accuracy}
\end{table}
\begin{table}[ht]
\centering
\begin{tabular}{lccccc}
  \hline
  Example & Time (TMB) & Speedup (TMB vs ADMB) \\
  \hline
  longlinreg &  11.3 &   0.9 \\
  mvrw &   0.3 &  97.9 \\
  nmix &   1.2 &  26.2 \\
  orange\_big &   5.3 &  51.3 \\
  sam &   3.1 &  60.8 \\
  sdv\_multi &  11.8 &  37.8 \\
  socatt &   1.6 &   6.9 \\
  spatial &   8.3 &   1.5 \\
  thetalog &   0.3 &  22.8 \\

  \hline
\end{tabular}
\caption{Timings for each example in seconds (Time) and speedup factor
  of TMB relative to ADMB (Speedup).}
\label{tab:speedup}
\end{table}
\begin{table}[ht]
\centering
\begin{tabular}{lrrrrrr}
  \hline  & sp chol & sp inv & AD init & AD sweep & GC & Other \\
  \hline    ar1\_4D & 71 & 22 &  1 &  1 &  3 &  2 \\
  ar1xar1 & 47 & 13 &  9 & 13 &  8 & 10 \\
  orange\_big &  3 &  4 & 20 & 57 &  6 & 11 \\
  sdv\_multi &  8 &  2 &  3 & 66 &  9 & 13 \\
  spatial &  1 &  0 & 17 & 71 &  2 &  9 \\
  \hline       
\end{tabular}
\caption{Percentage of time spent in the following
  (disjoint) parts of the algorithm: Sparse Cholesky factorization (sp
  chol),
  Inverse subset algorithm (sp inv), Initialization of tapes including
  automatic sparseness detection and tape optimization (AD init), AD forward and
  reverse mode sweeps (AD sweeps), \proglang{R} Garbage collection (GC) and
  remaining parts (Other). All
  examples were run with standard non-threaded \pkg{BLAS}. Note that the
  first two columns can be reduced by switching to a tuned \pkg{BLAS}. The
  middle two columns can, in certain cases, be reduced using parallel templates, while the
  final two columns are impossible to reduce for a single \proglang{R} instance.}
\label{tab:profile}
\end{table}

The case studies ``ar1\_4D'' and ``ar1xar1'' would be hard to
implement in \pkg{ADMB} because the sparsity would have to be manually
represented instead of automatically detected.  In addition, judging
from the speed comparisons presented below, \pkg{ADMB} would take a
long time to complete these cases.  Table~\ref{tab:compare_accuracy}
displays the difference of the results for \pkg{TMB} and \pkg{ADMB}
for all the case studies in Table~\ref{tab:description} (excluding
``ar1\_4D'' and ``ar1xar1'').  These differences are small enough to
be attributed to optimization termination criteria and numerical
floating point roundoff.  In addition, both packages were stable
w.r.t.~the choice of initial value.  Since these packages were coded
independently, this represents a validation of both package's software
implementation of maximum likelihood, the Laplace approximation, and
uncertainty computations.

For each of the case studies (excluding ``ar1\_4D'' and ``ar1xar1'')
Table~\ref{tab:speedup} displays the speedup which is defined as
execution time for \pkg{ADMB} divided by the execution time for
\pkg{TMB}.  In six out of the nine cases, the speedup is greater than
20; i.e., the new package is more than 20 times faster.  We note that
\pkg{ADMB} uses a special feature for models similar to the
``spatial'' case where the speedup is only 1.5.  The speedup is
greater than one except for the ``longlinreg'' case where it is 0.9.
This case does not have random effects, hence the main performance
gain is a result of improved algorithms for the Laplace approximation
presented in this paper and not merely a result of using a different
AD library.

\pkg{TMB} supplies an object with functions to evaluate the likelihood
function and gradient.  It is therefore easy to compare different
optimizers for solving the outer optimization problem.  We used this
feature to compare the \proglang{R} optimizers \code{optim} and
\code{nlminb}.  For the case studies in Table~\ref{tab:description},
the \code{nlminb} is more stable and faster than \code{optim}.  The
state-space assessment example ``sam'' was unable to run with
\code{optim} while no problems where encountered with \code{nlminb}.
For virtually all the case studies, the number of iterations required
for convergence was lower when using \code{nlminb}.

\begin{figure}
  \centering
  \includegraphics[width=8cm]{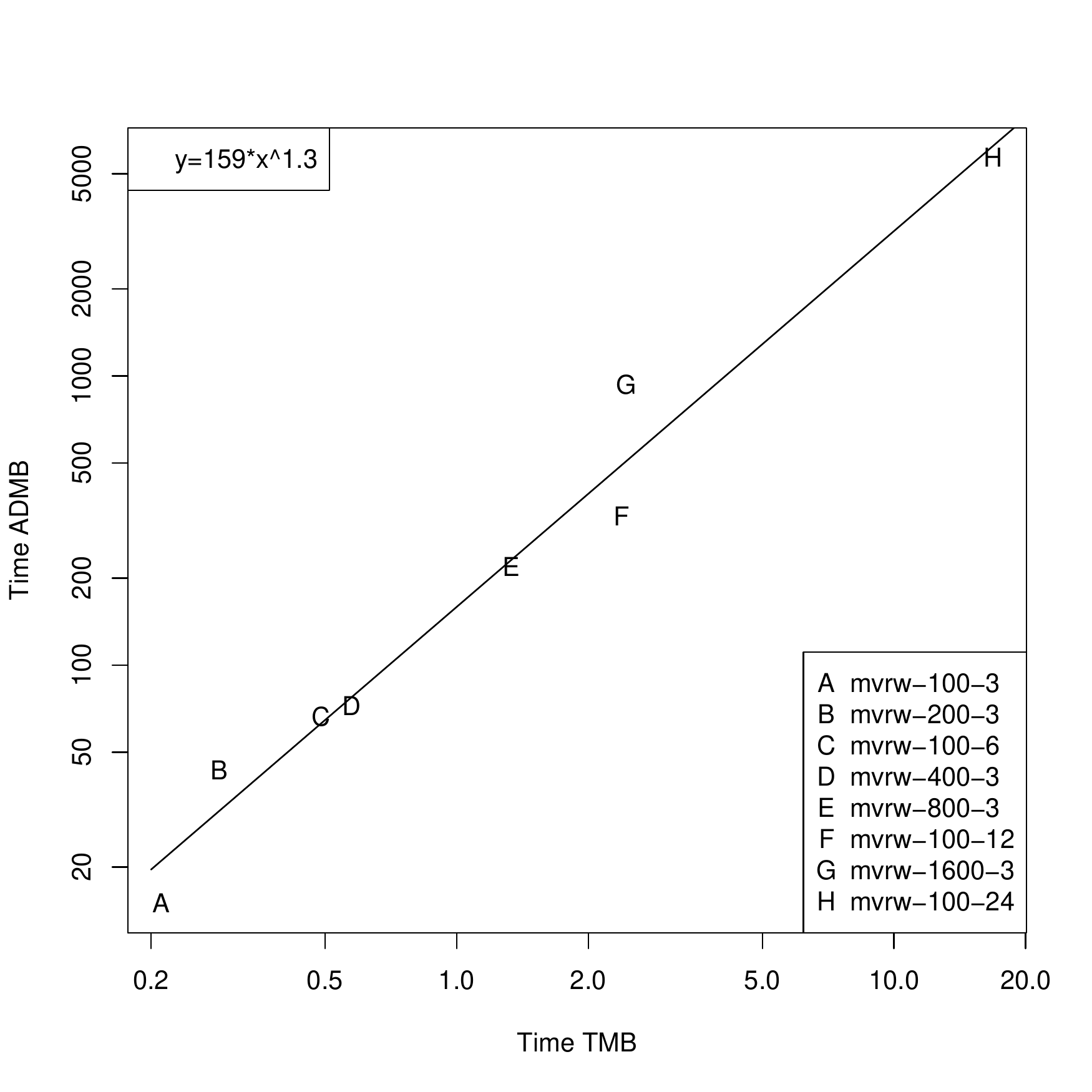}
  \caption{Run time comparison \pkg{ADMB} vs \pkg{TMB} of multivariate random walk example (mvrw) for varying number of time steps and block size.}
  \label{fig:mvrw_scaling}
\end{figure}
    
Most of the cases tested here have modest run times; to be specific,
on the order of seconds.  To compare performance for larger cases the
multivariate random walk ``mvrw'' was modified in two ways: 1) the
number of time steps was successively doubled 4 times; to be specific,
from 100 to 200, 400, 800, and 1600.  2) the size of the state vector
was also doubled 3 times; to be specific the dimension of $u_t$ was 3,
6, 12, and 24.  The execution time, in seconds $T$, for the two
packages is plotted in Figure~\ref{fig:mvrw_scaling}, and is a close
fit to the relation
\[
T_{\text{ADMB}} = 159 \times T_{\text{TMB}}^{1.3} \; .
\]  
While the parameters of this power-law relationship are
problem-specific, this illustrates that even greater speedups than
those reported in Table~\ref{tab:speedup} must be expected for larger
problems.  The case studies that took 5 or more seconds to complete
were profiled to identify their time consuming sections; see
Table~\ref{tab:profile}.  The studies fall in two categories.  One
category is the cases that spend over 50\% of the time in the sparse
Cholesky and inverse subset algorithms, a large portion of which is
spent in the \pkg{BLAS}.  This corresponds to the upper branch of
Figure~\ref{fig:package_graph}.  Performance for this category can be
improved by linking the \proglang{R} application to an optimized
\pkg{BLAS} library.  For example, the ``ar1\_4D'' case spends 93\% of
its time in these \pkg{BLAS} related routines.  Using the Intel MKL
parallel \pkg{BLAS} with 12 computational cores resulted in a factor
10 speedup for this case.  Amdal's law says that the maximum speedup
for this case is
\[
	6.8 = 1. / ( .93 / 12  + (1.0 - .93) ) \; .
\]
Amdal's law does not apply here because the MKL \pkg{BLAS} does other
speedups besides parallelization.

The other category is the cases that spend over 50\% of the time doing
AD calculations.  This corresponds to the lower branch of
Figure~\ref{fig:package_graph}.  For example, the ``sdv\_multi'' case
spends 66\% of the time doing AD sweeps.  We were able to speedup this
case using the techniques in Section~\ref{sec:parallel}.  The speedup
with 4 cores was a factor of 2.  Amdal's law says that the maximum
speedup for this case is
\[
1.98 = 1.0/(.66/4+(1-.66)) 
\]
which indicates that the AD parallelization was very efficient.  To
test this speedup for more cores we increased the size of the problem
from three state components at each time to ten.  For this case, using
10 cores, as opposed to 1 core, resulted in a 6 fold speedup.  (This
case is not present in Table~\ref{tab:profile} and we do not do an
Amdal's law calculation for it.)

The cheap gradient principle was checked for all the case studies.
The time to evaluate the Laplace approximation and its gradient
\code{obj\$gr} was measured to be smaller than 2.8 times the time to
evaluate just the Laplace approximation \code{obj\$fn}.  This is
within the theoretical upper bound of 4 calculated at the end of
Section~\ref{sec:inverse_subset}.  The factor was as low as 2.1 for
``sam'', the example with the highest number of parameters.

\section{Discussion}
\label{sec:Discussion}
This paper describes \pkg{TMB}, a fast and flexible \proglang{R}
package for fitting models both with and without random effects.  Its
design takes advantage of the following high performing and well
maintained software tools: \proglang{R}, \pkg{CppAD}, \pkg{Eigen},
\pkg{BLAS}, and \pkg{CHOLMOD}.  The collection of these existing tools
is supplemented with new code that implements the Laplace
approximation and its derivative. A key feature of \pkg{TMB} is that
the user do not have to write the code for the second order
derivatives that constitute the Hessian matrix, and hence provides an
``automatic" Laplace approximation.  This brings high performance
optimization of large nonlinear latent variable models to the large
community of \proglang{R} users.  A minimal effort is required to
switch a model already implemented in \proglang{R} to use \pkg{TMB}.
Post processing and plotting can remain unchanged.  This ease of use
will benefit applied statisticians who struggle with slow and unstable
optimizations, due to imprecise finite approximations of gradients.

The performance of \pkg{TMB} is compared to that of \pkg{ADMB}
\citep{ADModelBuilder2011}. In a recent comparative study among
general software tools for optimizing nonlinear models, \pkg{ADMB}
came out as the fastest \citep{bolker2013strategies}. In our case
studies, the estimates and their uncertainties were pratically
identical between \pkg{ADMB} and \pkg{TMB}. Since the two programs are
coded independently, this is a strong validation of both tools. In
terms of speed, their performances are similar for models without
random effects, however \pkg{TMB} is one to two orders of magnitude
faster for models with random effects.  This performance gain
increases as the models get larger.  These speed comparisons are for a
single core machine.

\pkg{TMB} obtains further speedup when multiple cores are available.
Parallel matrix computations are supported via the \pkg{BLAS} library.
The user specified template function can use parallel computations via
\pkg{OpenMP}.

An alternative use of this package is to evaluate, in \proglang{R},
any function written in \proglang{C++} as a ``user template'' (not
just negative log-likelihood functions).  Furthermore, the derivative
of this function is automatically available.  Although this only uses
a subset of \pkg{TMB}'s capabilities, it may be a common use, due to
the large number of applications in statistical computing that
requires fast function and derivative evaluation (\proglang{C++} is a
compiled language so its evaluations are faster).

Another tool that uses the Laplace approximation and sparse matrix
calculations (but not AD) is \pkg{INLA} \citep{rue2009approximate}.
\pkg{INLA} is known to be computationally efficient and it targets a
quite general class of models where the random effects are
Gauss-Markov random fields.  It would be able to handle some, but far
from all, of the case studies in Table \ref{tab:description}.  At the
least, the ``mvrw'' , ``ar1\_4D'' and ``ar1xar1'' cases.  It would be
difficult to implement the non-convex examples of Table
\ref{tab:description} in \pkg{INLA} because their likelihood functions
are very tough to differentiate by hand.

\pkg{INLA} uses quadrature to integrate w.r.t., and obtain a Bayesian
estimate of, the parameter vector $\theta$.  This computation time
scales exponentially in the number of parameters.  On the other
hand, it is trivial for \pkg{INLA} to perform the function evaluations
on the quadrature grid in parallel.  Using the parallel \proglang{R}
package, \pkg{TMB} could be applied to do the same thing; i.e.,
evaluate the quadrature points in parallel.

In conclusion, \pkg{TMB} provides a fast and general framework for
estimation in complex statistical models to the \proglang{R}
community. Its performance is superior to \pkg{ADMB}. \pkg{TMB} is
designed in a modular fashion using modern and high performing
software libraries, which ensures that new advances within any of
these can quickly be adopted in \pkg{TMB}, and that testing and
maintenance can be shared among many independent developers.
 
\bibliography{arxive}
\section{Supplementary Material}
\subsection{Parallel Templates}

\end{document}